\documentclass[a4paper,prd,noshowpacs,noshowkeys,twocolumn]{revtex4}
\usepackage[latin1]{inputenc}
\usepackage{amsmath}
\usepackage{amssymb}
\usepackage{mathrsfs}
\usepackage{graphicx}

\providecommand{\eq}[1]{\begin{equation} #1 \end{equation}}
\providecommand{\eqarr}[1]{\begin{eqnarray} #1 \end{eqnarray}}

\providecommand{\mt}[1]{\mbox{\tiny $#1$}}
\providecommand{\mn}[1]{\mbox{\normalsize $#1$}}
\providecommand{\ms}[1]{\mbox{\small $#1$}}
\providecommand{\mss}[1]{\mbox{\scriptsize $#1$}}
\def\Tr{\mathrm{Tr}}
\providecommand{\tp}{\mss{\mathsf{T}}}
\providecommand{\diag}{\mathrm{diag}}
\providecommand{\aver}[1]{\langle #1 \rangle}
\providecommand{\bs}[1]{\boldsymbol{#1}}
\providecommand{\hs}[1]{\hspace{#1}}

\providecommand{\br}{\mathbf{r}}
\providecommand{\bM}{\mathbf{M}}
\providecommand{\bbM}{\mathbb{M}}
\providecommand{\Mn}{\mathcal{M}_N}

\providecommand{\bt}{\mathbf{t}}
\def\sp{{\cdot}}

\font\bb=bbmss10 scaled 970
\def\id{\mbox{\bb 1}}
\begin{document}
%%%%%%%%%%%%%%%%%%%%%%%%%%%%%%%%%%%%%%%%%%%%%%%%%
\title{
Physical parameters and basis transformations in the Two-Higgs-Doublet model
}
\author{C.~C.~Nishi}
\email{ccnishi@ifi.unicamp.br}
\affiliation{
Instituto de Física ``Gleb Wataghin''\\
Universidade Estadual de Campinas, Unicamp\\
13083-970, Campinas, SP, Brasil
}
\affiliation{
Instituto de Física Teórica,
UNESP -- São Paulo State University\\
Rua Pamplona, 145,
01405-900 -- São Paulo, Brasil
}

%\date{\today}
%%%%%%%%%%%%%%%%%%%%%%%%%%%%%%%%%%%%%%%%%%%%%%%%%
\begin{abstract}
A direct connection between physical parameters of general Two-Higgs-Doublet
Model (2HDM) potentials after electroweak symmetry breaking (EWSB) and the
parameters that define the potentials before EWSB is established. These physical
parameters, such as the mass matrix of the neutral Higgs bosons, have well
defined transformation properties under basis transformations transposed to the
fields after EWSB. The relations are also explicitly written in a basis
covariant form. Violation of these relations may indicate models beyond 2HDMs.
In certain cases the whole potential can be defined in terms of the physical
parameters. The distinction between basis transformations and reparametrizations
is pointed out. Some physical implications are discussed.
\end{abstract}
%%%%%%%%%%%%%%%%%%%%%%%%%%%%%%%%%%%%%%%%%%%%%%%%%
%\keywords{ }
%\twocolumn
\maketitle
%%%%%%%%%%%%%%%%%%%%%%%%%%%%%%%%%%%%%%%%%%%%%%%%%
\section{Introduction}
\label{sec:intro}

The Standard Model (SM) relies on the Higgs mechanism to give masses to all
massive gauge bosons and elementary fermions. Such mechanism involves the
spontaneous symmetry breaking (SSB) of the electroweak $SU(2)_L\otimes U(1)_Y$
gauge group to the electromagnetic $U(1)_{\rm EM}$ as a scalar Higgs doublet
acquires a nonzero vacuum expectation value (VEV). Such scheme imposes
universality constraints on the couplings between fermions and gauge bosons,
establishes relations between the masses of the gauge bosons and fixes the
couplings between the physical Higgs and the fermions to be proportional to the
masses of the latter\,\cite{gomezbock:07}.

The scalar potential, constituted by only one Higgs doublet,
is also very restrictive since, from the knowledge of the electroweak (EW) VEV
and the physical Higgs mass, the Higgs trilinear and quartic coupling constants
are fixed at tree level. However, the physical Higgs boson, which is the only
scalar remnant of the EW Higgs mechanism, was not discovered yet. One then
resorts to indirect means to bound the physical Higgs mass, most of them relying
on the higher order perturbative behavior of the SM. Such bounds comes from,
\textit{e.g.}, the unitarity constraints of the scattering of gauge bosons, the
validity of the SM up to the Planck scale and the EW vacuum stability (see
Ref.\onlinecite{gomezbock:07} and references therein). Studies constraining the
Higss mass are very important to its search, in particular, in view of the
upcoming LHC experiment.

For models extending the electroweak symmetry breaking (EWSB) sector of the SM,
the relation of the model parameters before EWSB and the physical parameters
identified after EWSB may not be as minimal as in the SM. Specially for
N-Higgs-doublet extensions of the SM (NHDMs), the multiplicity of independent
parameters may be quite large due to the presence of a horizontal space, i.e.,
the space of identical gauge multiplets, in this case, $SU(2)_L$ doublets with
quantum numbers identical to the SM Higgs doublet.
The simplest Two-Higgs-Doublet Model (2HDM) has been extensively
studied
recently\,\cite{randall:2hdm,maniatis:1,BFS:2hdm,haber.davidson,haber.2,GH,
ivanov:mink,ivanov:mink2,ginzburg} as the effective scalar sector of the MSSM
requires two Higgs doublets for anomaly cancelation\,\cite{gomezbock:07,carena}.
Historically, the addition of one or more Higgs doublets were considered to
implement the spontaneous CP violation mechanism
(SCPV)\,\cite{lee:scpv,weinberg:scpv} as an alternative source of CP violation.	

Technical difficulties that arise when considering NHDMs are twofold: (i) more
than one local minimum (orbit), not necessary with the same symmetry breaking
pattern, might be present, even at tree level, and (ii) the reparametrization
freedom\,\cite{ginzburg} allowed by the presence of the horizontal space formed
by the N Higgs-doublets may masquerade the number of relevant independent
parameters and symmetry properties such as CP invariance. Difficulty (i)
includes the possibility of potentials with no remaining $U(1)_{\rm EM}$
symmetry after EWSB (charge breaking vacuum)\,\cite{sher,BFS:2hdm,BFSS:nhdm} and
it forces the stability of the vacuum to be a relevant issue at tree
level\,\cite{sher}. 

Item (ii) concerns the reparametrization transformations induced by basis
transformations (or horizontal transformations\,\cite{endnote0}) acting on the
identical N Higgs-doublets. Since all the doublets have the same gauge quantum
numbers with respect to the SM gauge group, there is no change in the physical
content of the theory if one rotates the fields in such
space\,\cite{GR,botellasilva:J,nhdm:cp}. Such possibility may masquerade the
number of relevant independent parameters in the theory. More crucially,
transforming real parameters into complex parameters (for complex multiplets),
CP invariant theories can be disguised as CP violating theories. This issue can
be solved in an objective way by noting that CP invariant theories remain CP
invariant through basis transformations but the corresponding CP symmetry
transformation also acts differently in different bases. As a practical way of
distinguishing the CP property of a theory one can resort to the use of
CP-odd basis invariants analogous to the Jarlskog invariant\,\cite{jarlskog} in
the quark sector of  the SM. In the context of NHDMs, numerous of such
invariants can be constructed\,\cite{botellasilva:J} but sufficient
conditions for CP invariance using a minimum number of invariants could be
formulated only for the potentials of
2HDMs\,\cite{botellasilva:J,GH,ivanov:cp,branco:05,nhdm:cp} and
3HDMs\,\cite{nhdm:cp}.

Recent advances in the study of 2HDMs include the result that at most two
local minima can be present whenever there is a discrete set of
minima in the orbit space\,\cite{ivanov:mink2,endnote-1}. Such result was
obtained by using a Minkowski structure that emerges naturally in the space of
the fields through suitable change of variables\,\cite{ivanov:mink,nhdm:cp}.
Although some controversy remain from numerical examples presenting more
than two local minima\,\cite{BFS:neutral}. Since the counting of the number of
local minima can be a very difficult task, the upper bound of the number of
minima is an important result.
It was also proved for 2HDMs that (a) charge breaking vacua can not coexist with
a neutral vacuum\,\cite{BFS:2hdm,ivanov:mink} and (b) spontaneously CP violating
vacua can not coexist with CP invariant
vacuum\,\cite{BFS:2hdm,ivanov:mink,ivanov:mink2}.
The result (b) can be extended in a weaker version to NHDMs: a spontaneously CP
violating extremum always lies above a CP invariant extremum if the latter
exists\,\cite{BFSS:nhdm}. The Minkowski structure can be also partially extended
to general NHDM potentials\,\cite{nhdm:V}.

Bearing these results in mind, the present article aims two goals concerning the
2HDM potential: (i) to extract all the physical parameters identifiable only
after EWSB and (ii) to study their properties under basis transformations.
The first goal involves having a more direct connection between the parameters
of the potential and the physical parameters after EWSB.
Since basis transformations are allowed before EWSB and they are usually
involved to reach the physical basis, it would be desirable to have a basis
covariant relation for the physical parameters, which leads to goal (ii).
%modified: 28.12.07
A systematic study of the physical parameters of 2HDMs, including the scalar
self-interactions and the interactions of scalars with fermions and gauge
bosons, was carried out in Ref.\,\onlinecite{haber.2}. The basis covariance,
however, was not extended to the fields after EWSB.

Another aspect of item (i) regards seeking a physical parametrization of the
2HDM potential by rewriting the parameters before EWSB in terms of the physical
parameters. As explained in the end of Ref.\,\onlinecite{nhdm:V}, what prevents
the utility of a parametrization depending on physical parameters, such as the
masses of the physical charged and neutral Higgs bosons, is the possibility of
the potential so defined possess another deeper minimum. The existence of at
most two local minima already ameliorate the situation. Such parametrization
also excludes by construction the potentials without nontrivial minima.
For the cases we know there is only one local minimum and hence it is also the
global one, such parametrization is unambiguous.
The question that remains is to know if such parametrization can cover all
possible 2HDM potential containing only one global minimum.

The ouline is as follows: in Sec.\,\ref{sec:PCH} we find the
mass matrix for neutral scalars in the basis where the mass matrix for charged
Higgs bosons is already diagonal. Relations between potential parameters and 
the mass matrices are found. In Sec.\,\ref{sec:gen}, covariant relations between
the mass matrices and the potential parameters are shown. In
Sec.\,\ref{sec:Pbasis} we show how to achieve the truly physical basis where all
the mass matrices are diagonal, pointing out the distinction between basis
transformations before and after EWSB. Finally, the results and physical
implications are discussed in Sec.\,\ref{sec:discussion}. Some possibly useful
material, including an alternative method to ensure a bounded below potential,
is presented in the appendices.

%%%%%%%%%%%%%%%%%%%%%%%%%%%%%%%%%%%%%%%%%%%%%%%%%
\section{Physical parameters in the PCH basis}
\label{sec:PCH}

A general 2HDM potential can be divided into its
quadratic and quartic parts as 
\eq{
\label{V:2+4}
V=V_2 + V_4\,.
}
The quadratic part is usually written as
\eq{
\label{V2:Y}
V_2=Y_{ab}\Phi^\dag_a\Phi_b\,,~~a,b=1,2,
}
where $Y$ is a hermitian matrix and $\Phi_a=(\phi_{a1},\phi_{a2})$ are the Higgs
doublets for which the notation $\phi_{a1}=\phi^{\mt{(+)}}_{a}$ and
$\phi_{a2}=\phi^{\mt{(0)}}_{a}$ is usually adopted when the vacuum preserves
the electromagnetic symmetry.
The quartic part can be conveniently written\,\cite{nhdm:cp} as
\eq{
\label{V4:Lamb}
V_4=\mn{\frac{1}{2}}\Lambda_{\mu\nu}r_\mu r_\nu\,,~~\mu,\nu=0,1,2,3,
}
where $\Lambda$ is a $4\times 4$ real symmetric matrix while
\eq{
\label{r.mu}
r_\mu=\mn{\frac{1}{2}}(\sigma_\mu)_{ab}\Phi_a^\dag\Phi_b~,
}
for $\mu=0,1,2,3$, are real quadratic combinations of the doublets. The matrix
$\sigma_0\equiv \id_2$ and $\sigma_i$ are the Pauli matrices. The quadratic
variables $r_\mu=(r_0,\br)$ are functionally free except for the future
lightcone constraint\,\cite{ivanov:mink,nhdm:V} 
\eq{
\label{LC}
r_0^2-\br^2\ge 0
\,,~~
r_0\ge 0\,.
}
The convention of summation over repeated indices is adopted with Euclidean
metric. For example, $r_\mu r_\mu=r^2_0 +\br^2$. The Minkowski metric will not
be used to avoid confusion and all indices will be written as lower indices,
differently of Refs.\,\cite{nhdm:V} and \cite{ivanov:mink}. 

Using the variables $r_\mu$, the quadratic part of the potential in
Eq.\,\eqref{V2:Y} can be cast into the form
\eq{
V_2= M_\mu r_\mu \,,
}
where $M_\mu$ has four independent components. The number of free parameters
contained in $Y$ and $M$ are the same and they are indeed related by
\eq{
Y=M_\mu\mn{\frac{1}{2}}\sigma_\mu\,
~\leftrightarrow~
M_\mu=\Tr[\sigma_\mu Y]\,.
}

We want to parametrize the potential in terms of physical quantities that are
defined after EWSB. Confining ourselves to neutral vacua, the first choice
of physical parameters will be the masses of physical particles, i.e., one
charged scalar and three neutral scalars.  In addition, more parameters such as
the EW vacuum expectation value (VEV), the mixing among the neutral scalars and
certain coupling constants will be necessary to completely parametrize the
potential that requires 11 essential parameters.

To extract the physical masses, we need the quadratic part of the potential
after EWSB that is induced by
\eq{
\label{shift:EWSB}
\Phi_a\rightarrow \aver{\Phi_a}+ \Phi_a\,,
}
where $\aver{\Phi_a}$ is the vacuum expectation value (VEV) of $\Phi_a$,
usually a c-number minimum of the potential in \eqref{V:2+4}. The extremum
equations are shown in appendix \ref{ap:min}.
With the shift of Eq.\,\eqref{shift:EWSB}, the quadratic part of the potential
can be written as\,\cite{nhdm:V}
\eq{
\label{V2:SSB}
V_2\big|_{\mt{\rm SSB}}=\Phi^{\dag}_a\aver{\mathbb{M}}_{ab}\Phi_b
+ \mn{\frac{1}{2}}\Lambda_{\mu\nu}s_\mu s_ \nu\,,
}
where
\eq{
\label{s.mu}
s_\mu=\mn{\frac{1}{2}}\aver{\Phi_a}^\dag(\sigma_\mu)_{ab} \Phi_b+h.c.,
}
and $\aver{\bbM}$ is the mass squared matrix for the charged Higgs bosons,
including the charged Goldstone. (The mass squared matrix will be denoted
simply as ``mass matrix'' from this point on.)
Such matrix can be calculated as\,\cite{nhdm:V}
\eq{
\label{M:charged}
\aver{\bbM}= Y+ \mn{\frac{1}{2}}\sigma_\mu\Lambda_{\mu\nu}\aver{r_\nu}
\,,
}
where
\eq{
\aver{r_\mu}=\mn{\frac{1}{2}}(\sigma_\mu)_{ab}\aver{\Phi_a}^\dag\aver{\Phi_b}\,.
}

In the Physical Charged Higgs (PCH) basis\,\cite{nhdm:V}, for a neutral
vacuum, the VEVs are simply 
\eq{
\label{vev:PCH}
\aver{\Phi_1}=
\begin{pmatrix}
0 \cr 0
\end{pmatrix}
,~~
\aver{\Phi_2}=
\begin{pmatrix}
0 \cr \frac{\mn{v}}{\sqrt{2}}
\end{pmatrix}
\,,
}
while the doublets after the shift \eqref{shift:EWSB} can be parametrized as
\eq{
\label{param:Phi}
\Phi_1=
\begin{pmatrix}
h^+ \cr \frac{1}{\sqrt{2}}(t_1-i t_2)
\end{pmatrix}
~, ~~
\Phi_2=
\begin{pmatrix}
G^+ \cr \frac{1}{\sqrt{2}}(-t_3+i G^0)
\end{pmatrix}
\,,
}
where $v=246\rm GeV$, is the electroweak VEV, $t_i$, $i=1,2,3$, are normalized
neutral scalar fields, $h^+$ is the physical charged Higgs and $G^+$ and $G^0$
are the charged and neutral Goldstone fields respectively. The Goldstone fields
$G^+$ and $G^0$ are absorbed by the longitudinal $W^+$ and $Z^0$ gauge bosons by
the Higgs mechanism.
For the VEV of Eq.\,\eqref{vev:PCH}, we have
\eq{
\label{vev:r.mu:PCH}
\aver{r_\mu}=\frac{v^2}{4}(1,0,0,-1)\,.
}

In the PCH basis, the mass matrix for the charged scalars can be written
\eq{
\label{bbM:PCH}
\aver{\bbM}=\diag(m^2,0)~,
}
where the null eigenvalue corresponds to the charged Goldstone.
We can divide the quadratic part of the potential of Eq.\,\eqref{V2:SSB} into
\eq{
V_2\big|_{\mt{\rm SSB}}
=V_2\big|_{\rm charged}+V_2\big|_{\rm neutral}
\,.
}
For the charged fields we have
\eq{
V_2\big|_{\rm charged}=m^2 h^+h^-
\,,
}
while
\eqarr{
V_2\big|_{\rm neutral}&=&
\mn{\frac{1}{2}}m^2(t^2_1+t^2_2)+
\mn{\frac{1}{2}}\Lambda_{\mu\nu}s_\mu s_ \nu\,,
\\
\label{Mn:ti}
&=& \mn{\frac{1}{2}}t_i(\Mn)_{ij}t_j
\,,
}
where, using the parametrization of Eq.\,\eqref{param:Phi},
\eq{
\label{ti:PCH}
s_0= -\frac{v}{2}t_3, ~~s_i=\frac{v}{2}t_i\,.
}
The $3\times 3$ matrix $\Mn$ is the mass matrix for the physical neutral
scalars given by
\eqarr{
\label{M:neutral}
\Mn&=&
m^2\diag(1,1,0)
+\frac{v^2}{4}\tilde{\Lambda}
\cr
&&~
+\frac{v^2}{4}
\begin{pmatrix}
0 & 0 & -\Lambda_{01}\cr
0 & 0 & -\Lambda_{02}\cr
-\Lambda_{01} & -\Lambda_{02} & \Lambda_{00}-2\Lambda_{03}
\end{pmatrix}
\,,
}
where $\tilde{\Lambda}=\{\Lambda_{ij}\}$, $i,j=1,2,3$.
The physical neutral scalars will be orthogonal combinations of $t_i$, defined
by the diagonalization of $\Mn$ in Eq.\,\eqref{M:neutral}.
% modified: 28.12.07
The mass matrix $\Mn$ in the PCH basis can be also found in Eqs.\,(24) and (41)
of Ref.\,\onlinecite{haber.2} in a different notation.

From Eqs.\,\eqref{M:charged}, \eqref{vev:r.mu:PCH} and \eqref{M:neutral} we can
find the following relation between $Y$ and $\Mn$,
\eq{
\label{param:Y}
Y=\mn{\frac{1}{2}}
\begin{pmatrix}
2Y_{11} & (\Mn)_{13}-i(\Mn)_{23}\cr
(\Mn)_{13}+i(\Mn)_{23} & -(\Mn)_{33}
\end{pmatrix}
\,.
}
Except for $Y_{11}$, all elements of $Y$ are directly related to $\Mn$.

Hence, we can completely parametrize the potential in the PCH basis in terms of
the set of 12 parameters
\eq{
\label{params}
\{v,m,\Lambda_{00},\Mn,\bs{\Lambda_0}\}
\,,
}
where $\bs{\Lambda_0}=\{\Lambda_{0i}\}$, $i=1,2,3$.
For fixed values for the set in Eq.\,\eqref{params}, we obtain from
Eq.\,\eqref{M:neutral} the rest of $\Lambda_{\mu\nu}$ by
\eqarr{
\label{La12}
\Lambda_{ij}&=&\frac{4}{v^2}[(\Mn)_{ij}-m^2\delta_{ij}],~~i,j=1,2\,,\\
\label{La13}
\Lambda_{i3}&=&\Lambda_{0i}+\frac{4}{v^2}(\Mn)_{i3},~~i=1,2\,,\\
\label{La33}
\Lambda_{33}&=&-\Lambda_{00}+2\Lambda_{03}+\frac{4}{v^2}(\Mn)_{33}\,.
}
The quadratic parameter $Y_{11}$ depends on more parameters other than $\Mn$
as
\eq{
\label{param:Y11}
Y_{11}=m^2-\frac{v^2}{4}(\Lambda_{00}-\Lambda_{03})+\mn{\frac{1}{2}}(\Mn)_{33}
\,.
}

There are 12 free parameters. Among these, 11 are essential and can not be
eliminated by reparametrization\,\cite{nhdm:cp}. Nevertheless, one parameter can
be removed by a remaining $U(1)$ reparametrization freedom due to 
\eq{
\label{reparam:Phi1}
\Phi_1 \rightarrow e^{i\theta} \Phi_1\,,~~
\Phi_2 \rightarrow \Phi_2 \,.
}
Since
\eq{
h^+ \rightarrow e^{i\theta} h^+
}
represents the electromagnetic $U(1)_{\rm EM}$ invariance of the potential, the
transformation of Eq.\,\eqref{reparam:Phi1} amounts effectively to a
$SO(2)$ rotation in the $t_1,t_2$ fields. Notice that since $\aver{\Phi_1}=0$,
the transformation of Eq.\,\eqref{reparam:Phi1} does not affect the VEVs.
Choosing appropriately $\theta$ in Eq.\,\eqref{reparam:Phi1}, one can set one of
the following parameters to zero: $(\Mn)_{12},(\Mn)_{13},(\Mn)_{23}$. In
particular, choosing $(\Mn)_{23}=0$, we obtain a real symmetric $Y$ and we
constrain $\Lambda_{23}=\Lambda_{03}$. For any choice the overall number of free
independent parameters should be 11. Of course, different choices, such as
$\Lambda_{01}=0$ or $\Lambda_{02}=0$, could be alternatively chosen.
Obviously, once a choice is made, one can not set more than one parameter to
zero.

To ensure the vacuum in Eq.\,\eqref{param:Phi} to be a local minimum, it is
sufficient to pick positive values for $m^2$ (the mass squared of $h^+$) and
for the three eigenvalues of $\Mn$ (the masses squared of $t_i$). Such
requirements guarantee that the second derivative of the potential around the
extremum is positive semidefinite.

There remains the question of boundedness for the potential defined with
physical parameters \eqref{params}. Firstly, we have to choose $\Lambda_{00}\ge
0$ because taking $r_0\rightarrow \infty$ but $|\br|$ finite in
Eq.\,\eqref{V4:Lamb} would make $V_4$ acquire negative values if $\Lambda_{00}<
0$. Moreover, the following statement can be proved:
\begin{quote}
For a potential $V(r)$ defined as Eq.\,\eqref{V:2+4}, satisfying
$\Lambda_{00}+\lambda_i>0$ for all $\lambda_i$, $i=1,2,3$, eigenvalues of
$\tilde{\Lambda}$, it is always possible to obtain $\Lambda_{\mu\nu}r_\mu
r_\nu>0$, for all $r_\mu$ satisfying Eq.\,\eqref{LC}, by making the substitution
\eq{
\label{BB:L0}
\bs{\Lambda_0}\rightarrow c\bs{\Lambda_0}\,,~~c>0\,,
}
with appropriately small $c$.
\end{quote}
The proof is shown in appendix~\ref{ap:BBcond}

The only problem that could make such physical parametrization not viable is the
possibility that the potential defined for a given set \eqref{params} possess
another minimum that lies deeper than the one defined in \eqref{vev:PCH}. This
possibility is real and numerical examples can be quickly deviced. The problem
is not so severe because at most two distinct local minima are possible for
bounded below potentials containing two Higgs doublets\,\cite{ivanov:mink2}. 
Although, in Ref.\,\onlinecite{BFS:neutral}, some numerical examples of 2HDM
potentials with more than two minima were apparently
devised\,\cite{BFS:neutral}.
On the other hand, this parametrization excludes pontentials without
nontrivial minima by construction.

%%%%%%%%%%%%%%%%%%%%%%%%%%%%%%%%%%%%%%%%%%%%%%%%%%%%
\section{Physical parameters in an arbitrary basis}
\label{sec:gen}

For a general potential \eqref{V:2+4}, the vacuum expectation value (VEV) will
not be in the form of Eq.\,\eqref{vev:PCH}. Nonetheless, we can
always parametrize a neutral vacuum as
\eq{
\label{vev:gen}
\aver{\Phi_1}=
\frac{v}{\sqrt{2}}
\begin{pmatrix}
0 \cr \cos\mn{\frac{\ms{\theta_v}}{\ms{2}}}
\end{pmatrix}
~, ~~
\aver{\Phi_2}=
\frac{v}{\sqrt{2}}
\begin{pmatrix}
0 \cr e^{i\xi}\sin\mn{\frac{\ms{\theta_v}}{\ms{2}}}
\end{pmatrix}
\,.
}
The VEV in Eq.\,\eqref{vev:r.mu:PCH} corresponds to $\theta_v=\pi$ and
$\xi=\pi$. In the MSSM, the angle $\theta_v$ corresponds to $2\beta$.

To explicit the structure of the horizontal space where basis transformations
act, it is more convenient to define\,\cite{nhdm:V}
\eq{
\label{uw}
u\equiv (\phi_{11},\phi_{21})^{\tp}~,~~
w\equiv (\phi_{12},\phi_{22})^{\tp}~.
}
We can rewrite Eq.\,\eqref{vev:gen} as
\eq{
\label{<u,w>}
\aver{u}=(0,0)^{\tp},~~
\aver{w}=\frac{v}{\sqrt{2}}
(\cos\!\mn{\frac{\ms{\theta_v}}{\ms{2}}},
\sin\!\mn{\frac{\ms{\theta_v}}{\ms{2}}}\,e^{i\xi} )^{\tp}
\,.
}
More generally, we can rewrite
\eq{
\label{vev:w:gen}
\aver{w}= \frac{v}{\sqrt{2}}U_v e_2\,,
}
where $U_v$ is a unitary matrix in $SU(2)_H$ and $e_a$, $a=1,2$, are the
canonical vectors defined by $(e_a)_b=\delta_{ab}$.
In terms of $r_\mu$ we get
\eq{
\label{vev:r.mu:gen}
\aver{r_\mu}=\frac{v^2}{4}
(1,\cos\xi\sin\theta_v,\sin\xi\sin\theta_v,\cos\theta_v)
=\frac{v^2}{4}R_{\mu\nu}n_\nu\,,
}
where $n_\mu=(1,0,0,-1)$ and $R_{\mu\nu}$ can be related to $U_v$ by
\eq{
\label{R.munu}
R_{\mu\nu}(U_v)\equiv \mn{\frac{1}{2}}\Tr[U^\dag_v\sigma_\mu U_v\sigma_\nu]
=
\begin{pmatrix}
1 & \bs{0} \cr
\bs{0} & \tilde{R}
\end{pmatrix}
\,,
}
with $\tilde{R}=\{R_{ij}\}$, $i,j=1,2,3$, being a rotation matrix in $SO(3)_H$.

We can rewrite the quadratic part of the potential in Eq.\,\eqref{V2:SSB} using
$u$ and $w$ of Eq.\,\eqref{uw}, and their respective VEVs in
Eq.\,\eqref{<u,w>}, 
\eq{
\label{V2:SSB:uw}
V_2\big|_{\mt{\rm SSB}}=
u^\dag\aver{\bbM}u
+ w^\dag\aver{\bbM}w
+ \mn{\frac{1}{2}}\Lambda_{\mu\nu}s_\mu s_\nu
\,,
}
where
\eq{
\label{s.mu:w}
s_\mu=\mn{\frac{1}{2}}\aver{w}^\dag\sigma_\mu w + h.c.
}
The first term of Eq.\,\eqref{V2:SSB:uw} corresponds to the mass term of
the charged scalars, one physical and one Goldstone, in an arbitrary basis.
The respective mass matrix is $\aver{\bbM}$, which is defined by
Eq.\,\eqref{M:charged}.

The relation between $\aver{\bbM}$ in an arbitrary basis and its diagonal
form\,\eqref{bbM:PCH} in the PCH basis is given by
\eq{
\label{bbM->diag}
U_v^\dag\aver{\bbM}U_v=\diag(m^2,0)\,.
}
We can then reach the PCH basis by the substitutions
\eq{
\label{Phi->PCH}
\Phi_a = (U_v)_{ab}\Phi'_b\,,~~
\aver{\Phi_a} = (U_v)_{ab}\aver{\Phi'_b}\,,
}
or equivalently
\eq{
\label{uw->PCH}
w=U_v w', ~~u=U_v u'\,,
}
with the same substitutions valid for their respective VEVs.

%%% trecho retirado

Since the basis for which $\{Y,\Lambda,\aver{r}\}$ is defined is
completely arbitrary, the covariance of $\aver{\bbM}$ is valid between any basis
and not only with respect to the PCH basis. (A detailed account of the basis
covariance of $\aver{\bbM}$ is given in appendix \ref{ap:covariance:M}.)
Indeed, we can write\,\cite{nhdm:V}
\eq{
\label{bbM->w}
\aver{\bbM}=m^2\big[\id_2-\aver{\hat{w}}\aver{\hat{w}}^\dag\big]\,,
}
where $\aver{\hat{w}}=\aver{w}/|\aver{w}|$.

We can try to extend the basis covariance for the mass matrix for the physical
neutral scalars. We keep the notation $\Mn$ to denote such mass matrix.
Obviously, the second and third term of
Eq.\,\eqref{V2:SSB:uw} is covariant by basis transformations for $w$, such 
as the transformation \eqref{uw->PCH} with arbitrary $U_v$. The question,
however, is if we can find appropriate fields $t_i$, with suitable
transformation properties, that renders $\Mn$ covariant by some basis
transformation, keeping Eq.\,\eqref{Mn:ti} form invariant. We obviously want to
recover Eq.\,\eqref{M:neutral} for $\Mn$ and Eq.\,\eqref{ti:PCH} for $t_i$ in
the PCH basis.

The immediate extension of Eq.\,\eqref{param:Phi} to define $t_i$ in any basis
is flawed because a basis transformation over $w$ in Eq.\,\eqref{param:Phi}
would mix $t_i$ with the neutral Goldstone $G^0$. In other words, with $w'$ in
the PCH basis given by 
\eq{
w'=
\frac{1}{\sqrt{2}}
\begin{pmatrix}
t'_1 - i t'_2 \cr
-t'_3+iG^0
\end{pmatrix}
\,,
}
we can not define 
\eq{
w=
\frac{1}{\sqrt{2}}
\begin{pmatrix}
t_1 - it_2 \cr
-t_3+iG^0
\end{pmatrix}
}
because, in general, 
\eq{
w\neq U_vw'\,,
}
such as for $U_v=e^{i\theta \sigma_1/2}$.

The solution is to promote Eq.\,\eqref{ti:PCH} to define the real fields
$t_i$ as
\eq{
\label{ti->si}
t_i\equiv \frac{2}{v}s_i\,,~~i=1,2,3\,.
}
The definition of Eq.\,\eqref{ti->si} ensures that $t_i$ would transform as
vectors under $SO(3)_H$ when $SU(2)_H$ transformations are applied to $w$ and
$\aver{w}$ in the definition \eqref{s.mu:w} of $s_i$.
The $s_0$ component depends on $t_i$ by the basis invariant relation
\eq{
\label{s0->ti}
s_0=\frac{v}{2}\aver{\hat{\br}}\sp\,\bt\,,
}
where $\aver{\hat{\br}}$ is the unit vector in the direction of $\aver{\br}$.
The relation \eqref{s0->ti} is proved in appendix \ref{ap:s0->ti}.

To write the second and third terms of Eq.\,\eqref{V2:SSB:uw} in terms of $t_i$
it is necessary to find the parametrization of $w$ in terms of $t_i$ and $G^0$.
The desired covariant relation is 
\eq{
\label{w->ti}
w=\frac{1}{\sqrt{2}}(iG^0\id_2+t_i\sigma_i)\aver{\hat{w}}\,,
}
where $\aver{\hat{w}}=\aver{w}/|\aver{w}|$.
One can confirm that Eq.\,\eqref{w->ti} satisfies Eq.\,\eqref{ti->si} using
Eq.\,\eqref{s.mu}.
The covariance can be also checked,
\eq{
w'=Uw=\frac{1}{\sqrt{2}}(iG^0\id_2+t'_i\sigma_i)\aver{\hat{w}'}\,,
}
where
\eq{
t'_i=R_{ij}t_j\,,
}
and $R_{ij}$ is related to $U$ by a relation similar to Eq.\eqref{R.munu}.
The dependence on $G^0$ is fixed by imposing that Eq.\,\eqref{w->ti}
reduces to Eq.\,\eqref{param:Phi} in the PCH basis.
% modified: 28.12.07
Notice Eq.\,\eqref{w->ti} differs from Eq.\,(46) of
Ref.\,\onlinecite{haber.2} as the fields $t_i$ transform as vectors under
$SO(3)_H$.

We can thus rewrite the second term of Eq.\,\eqref{V2:SSB:uw} in
terms of $t_i$ using Eqs.\,\eqref{bbM->w} and \eqref{w->ti} as
\eq{
w^\dag\aver{\bbM}w=\frac{m^2}{2}[\bt^2-(\aver{\hat{\br}}\sp\bt)^2]
\,.
}
The third term of Eq.\,\eqref{V2:SSB:uw} can be also easily rewritten in terms
of $t_i$ by using Eq.\,\eqref{ti->si}.
The sum of the second and the third terms of Eq.\,\eqref{V2:SSB:uw} defines
the mass matrix for the physical neutral scalars by
\eq{
w^\dag\aver{\bbM}w+\mn{\frac{1}{2}}\Lambda_{\mu\nu}s_\mu s_\nu
=\mn{\frac{1}{2}}(\Mn)_{ij}t_it_j
\,,
}
giving the basis covariant relation
\eqarr{
\label{Mn:gen}
\Mn&=&
m^2\Big[\id_3- \frac{\aver{\br}\aver{\br}^{\tp}}{|\aver{\br}|^2}\Big]
+|\aver{\br}|\Big[\tilde{\Lambda}+
\Lambda_{00}\frac{\aver{\br}\aver{\br}^{\tp}}{|\aver{\br}|^2}\Big]
~~\cr&&~
+\aver{\br}\bs{\Lambda_0}^{\tp}+\bs{\Lambda_0}\aver{\br}^{\tp}
\,,
}
where $|\aver{\br}|=v^2/4$.
Equation \eqref{Mn:gen} is the generalization of Eq.\,\eqref{M:neutral} for an
arbitrary basis.

Equation \eqref{Mn:gen} illustrates two points: (i) the whole potential after
EWSB can be completely defined in terms of the set
\eq{
\label{params:gen}
\{m^2,\Lambda_{00},\bs{\Lambda_0},\aver{\br},\Mn\}
\,,
} 
in an arbitrary basis, since $\tilde{\Lambda}$ can be written in terms of the
set.
Moreover, the objects in the set have the same transformation properties under
the reparametrization group $SU(2)_H$ as the set
$\{M_0,\Lambda_{00},\bs{\Lambda_0},\bM,\tilde{\Lambda}\}$ that defines the
potential before EWSB\,\cite{nhdm:cp}: two scalars\,\cite{endnote1}, two vectors
and one rank-2 tensor. (ii) We can define a Physical Neutral Higgs (PNH) basis,
in contrast to the PCH basis, being the basis where $\Mn$ is diagonal. In
general, this basis will coincide neither with the PCH basis nor with the basis
with diagonal $\tilde{\Lambda}$ (the canonical CP basis in
Ref.\,\cite{nhdm:cp}).

The relations \eqref{param:Y} and \eqref{param:Y11} can be written
in the basis covariant form
\eqarr{
\label{Y:gen}
Y&=&
\mn{\frac{1}{2}}[m^2-\aver{r_0}\Lambda_{00}-\bs{\Lambda_0}\sp\aver{\br}]
\big(\id_2-\sigma_i\aver{\hat{r}_i}\big)\cr
&&\,
-\,\mn{\frac{1}{2}}\sigma_i(\Mn)_{ij}\aver{\hat{r}_j}
\,.
}
We made use of the relation
\eq{
\id_2-\aver{\hat{w}}\aver{\hat{w}}^\dag=
\mn{\frac{1}{2}}\big[\id_2-\sigma_i\aver{\hat{r}_i}\big]
\,.
}
One can check Eq.\,\eqref{Y:gen} reduces to Eqs.\,\eqref{param:Y} and
\eqref{param:Y11} in the PCH basis.
It is also important to remark that $Y$ in Eq.\,\eqref{Y:gen} is independent on
the particular VEV. For a different minimum of the same potential (or extremum
if we do not require positive definite $\Mn$ and $m^2$), $\Mn,m^2$ and
$\aver{\br}$ differ in such a way that $Y$ is the same.
In addition, we can write the $(ij)=(33)$ component of Eq.\,\eqref{param:Y} in
the following basis covariant form
\eq{
\label{wYw->rMr}
\aver{\hat{w}}^\dag Y\aver{\hat{w}}=
- \mn{\frac{1}{2}}\aver{\hat{r}_i}(\Mn)_{ij}\aver{\hat{r}_j}\,.
}

To obtain the interaction terms\,\cite{nhdm:V}
\eqarr{
\label{V3:SSB}
V_3\big|_{\mt{\rm SSB}}&=&\Lambda_{\mu\nu}s_\mu r_\nu\,, \\
\label{V4:SSB}
V_4\big|_{\mt{\rm SSB}}&=&\mn{\frac{1}{2}}\Lambda_{\mu\nu}r_\mu r_\nu\,,
}
in terms of the real fields $t_i$, we should calculate $r_\mu$
in Eq.\,\eqref{r.mu} using Eq.\,\eqref{ti->si}.
Splitting 
\eq{
r_\mu=x_\mu+y_\mu\,,
}
where
\eqarr{
\label{x.mu}
x_\mu&=&\mn{\frac{1}{2}}u^\dag\sigma_\mu u\,,\\
y_\mu&=&\mn{\frac{1}{2}}w^\dag\sigma_\mu w\,,
}
we obtain the following covariant relations for $y$,
\eqarr{
\label{y0->t}
y_0&=&\mn{\frac{1}{4}}[(G^0)^2+\bt^2]\,,\\
\label{yi->t}
y_i&=&\mn{\frac{1}{4}}[(G^0)^2-\bt^2]\aver{\hat{r}_i}
+\mn{\frac{1}{2}}(\aver{\hat{\br}}\sp\bt)t_i
+\mn{\frac{1}{2}}G^0 (\bt{\times}\aver{\hat{\br}})_i
\,.\hs{1.5em}
}
Since each component of $u=(u_1,u_2)^{\tp}$ is a combination of the
physical charged Higgs $h^+$ and the charged Goldstone $G^+$, there is no need
to write them in terms of real fields. The expression \eqref{x.mu} can be kept
as the covariant relation.

% modified: 28.12.07
A last observation about Eq.\,\eqref{yi->t} concerns the transformation
properties of $G^0$ under refections, i.e., CP symmetry. To keep the
transformation properties of the last term of Eq.\,\eqref{yi->t} to be the same
as the preceding terms we conclude that $G^0$ should be a pseudoscalar (scalar
under $SO(3)_H$ and changing sign under reflection or CP) and consequently
CP-odd irrespective of the CP properties of the potential. (See appendix D of
Ref.\,\onlinecite{haber.2}.)

%%%%%%%%%%%%%%%%%%%%%%%%%%%%%%%%%%%%%%%
\section{Physical basis (P-Basis)}
\label{sec:Pbasis}

The Physical basis (P-basis) should be defined as the basis where all the fields
possess definite masses. From Sec.\,\ref{sec:PCH}, we conclude that the mass
matrix for physical neutral scalars ($\Mn$) in the PCH basis will be not
diagonal in general. From Sec.\,\ref{sec:gen}, the basis where $\Mn$ is diagonal
(PNH basis) would mix the physical charged Higgs $h^+$ with the charged
Goldstone $G^+$. Thus, neither the PCH basis nor the PNH basis coincide with the
Physical basis.

To achieve the P-basis, we need independent basis transformations on the upper
($u$) and lower components ($w$) of the doublets, i.e.,
\eq{
\label{uw:trans:gen}
u\rightarrow U_u u\,,~~w\rightarrow U_w w\,,
}
where $U_u,U_w$ are different transformations in $SU(2)$. The transformations in
Eq.\,\eqref{uw:trans:gen} are legitimate basis transformations only after EWSB
since they still preserve the EM symmetry, for a neutral vacuum, but do not
preserve the $SU(2)_L\otimes U(1)_Y$ gauge structure of the doublets, except for
$U_u=U_w$. The general group of basis transformations generated by
Eq.\,\eqref{uw:trans:gen} is $SU(2)\otimes SU(2)$ instead of $SU(2)_H$ valid
before EWSB. 

The P-basis can be achieved either from the PCH basis or from the PNH basis. The
latter choice is more convenient.
Let us choose the PNH basis for which
\eq{
\Mn=\diag(m^2_1,m^2_2,m^2_3)\,.
}
The VEV $\aver{w}$ will be in the general form of Eq.\,\eqref{vev:w:gen},
different from the PCH basis. The respective $\aver{r_\mu}$ would be
parametrized in the form of Eq.\,\eqref{vev:r.mu:gen}.

In the PNH basis, the fields $t_i$ and $G^0$, contained in $w$, already have
definite masses. The components of $u$, however, are combinations of the
physical charged fields $h^+$ and $G^+$. The relation between $u$ and the
physical fields is given by
\eq{
\label{u->u'}
u=U_v u'\,,
}
where $u'$ refers to $u$ in the PCH basis,
\eq{
u'=(h^+,G^+)^{\tp}\,.
}
In a basis invariant form, we know the component of $u$ parallel to $\aver{w}$
is the charged Goldstone,
\eq{
\aver{\hat{w}}^\dag u=G^+\,.
}
The orthogonal direction contains the physical $h^+$.

Obviously, the quadratic part of the potential after EWSB will be
\eq{
V_2\big|_{\mt{\rm SSB}}=m^2 h^+h^- 
+ \mn{\frac{1}{2}}m^2_i t^{2}_i\,.
}
The remaining task to completely define the potential after EWSB is to write
the interaction terms in Eqs.\,\eqref{V3:SSB} and \eqref{V4:SSB} in terms of
$\{t_i,G^0,h^+,G^+\}$. The sole dependence of those interaction terms on $u$
comes from $x_\mu$ in Eq.\,\eqref{x.mu}. The component $x_0$ is basis
independent and can be readily written 
\eq{
x_0=\mn{\frac{1}{2}}(G^-G^+ + h^-h^+)\,.
}
The spatial components can be written
\eq{
\label{x->x'}
x_i=(\tilde{R}_{v})_{ij}x'_j\,,
}
where $(\tilde{R}_v)_{ij}=\Tr[U_v^\dag\sigma_iU_v\sigma_j]$ and
\eqarr{
x'_1&=&\mn{\frac{1}{2}}(h^-G^+ + h^+G^-)\,,\\
x'_2&=&\mn{\frac{-i}{2}}(h^-G^+ - h^+G^-)\,,\\
x'_3&=&\mn{\frac{1}{2}}(G^-G^+ - h^-h^+)\,.
}
The variables $s_\mu$ can be written in terms of $t_i$ using
Eqs.\,\eqref{ti->si} and \eqref{s0->ti} while the variables $y_\mu$ are defined
in Eqs.\,\eqref{y0->t} and \eqref{yi->t}.

%%%%%%%%%%%%%%%%%%%%%%%%%%%%%%%%%%%%%%%
\section{Discussions}
\label{sec:discussion}

Equation\,\eqref{wYw->rMr} relates the depth of the potential in the minimum
with the mass matrix of the physical neutral scalars. We can obtain bounds on
the depth of the potential from the relations
\eq{
\label{<V>}
V(\aver{r})=\mn{\frac{1}{2}}V_2(\aver{r})= -V_4(\aver{r})\,,
}
where $\aver{r}$ represents an extremum while $V_2$ and
$V_4$ refer respectively to the quadratic and quartic part of
the potential before EWSB, defined in Eqs.\,\eqref{V2:Y} and \eqref{V4:Lamb},
evaluated in the extremum.
The first equality of Eq.\,\eqref{<V>} can be written using
Eq.\,\eqref{wYw->rMr} as
\eq{
V(\aver{r})=-\mn{\frac{1}{2}}
\aver{r_0}\aver{\hat{r}_i}^{\tp}(\Mn)_{ij}\aver{\hat{r}_j}\,.
}
From the relation above, we can deduce the following bounds for the depth of a
minimum $\aver{r}$,
\eq{
-\mn{\frac{1}{2}}\aver{r_0}m^2_3 \le V(\aver{r})\le
-\mn{\frac{1}{2}}\aver{r_0}m^2_1\,,
}
where $m^2_3$ and $m^2_1$ are respectively the greatest and the least eigenvalue
of $\Mn$. We can conclude that a minimum will be deeper if the respective masses
for the neutral scalars and the value of $v$ are greater.

A different physical bound can be extracted from condition
\eqref{BB:Lt} necessary for bounded below potentials and from the positive
definiteness of $\Mn$. From Eq.\,\eqref{Mn:gen} and 
$e_{\perp}^{\tp}(\tilde{\Lambda}+\Lambda_{00}\id)e_{\perp}>0$ we
arrive at
\eq{
m^2-m^2_3< \frac{v^2}{4} \Lambda_{00}\,,
}
where $e_{\perp}$ is any unit vector orthogonal to $\aver{\hat{\br}}$.
The last inequality means the mass of the charged Higgs can not be
arbitrarily large compared to the masses of the neutral scalars.

In Sec.\,\ref{sec:gen} we have found the set of Eq.\,\eqref{params:gen} could
be chosen as the physical parameters that define the 2HDM potential with a
nontrivial vacuum.
Among the elements of the set, it is clear that the masses are physical
observables.
On the other hand, the connection of the coupling constants and mixing matrices
appearing in the interaction terms with physical observables is not
direct. For example, devising scattering observables to extract the three
parameters composing $\bs{\Lambda_0}$, present in $V_3$ and $V_4$, does not
seem a straightfoward task. The form of $V_3$ in terms of the physical fields,
given in appendix \ref{ap:V34}, reinforce such difficulty. 
% 28.12.07{
The explicit form of $V_3$ and $V_4$ in the Physical basis can be also found in
Eqs.\,(57)--(60) of Ref.\,\onlinecite{haber.2}, although the dependence on the
mass matrix of the neutral scalars are not explicitly shown.
% 28.12.07}
An attempt to extract the observable parameters in the 2HDM, aiming to
identify the presence of discrete symmetries through measurements, was made in
Ref.\,\onlinecite{lavoura:discrete}.
Nevertheless, separating the set of Eq.\,\eqref{params:gen} and finding the
relation of other parameters with the set is important to establish the
number of independent parameters possible. The violation of any relation between
parameters would indicate a model with a scalar sector distinct of the 2HDM. 
%modified: 31.12.07<
These relations should be constrained by experimental data and studies of
the bounds on the mass of the physical charged
Higgs\,\cite{barenboim:07,krawczyk:07} or of the decay width of the
physical Higgs bosons\,\cite{randall:2hdm,trott} already exist in the
literature.
%modified: 31.12.07>
Of course, higher order effects, such as the exchange of quarks, would modify
these tree level relations.
%modified: 31.12.07<
The number of minima may be also modified when higher order contributions are
taken into account. The existence of at most two minima, for example, may not
be true beyond tree level\,\cite{endnote:ivanov}.
%modified: 31.12.07>

Another aspect of the identification of physical parameters concerns the
remaining reparametrization freedom such as the one in
Eq.\,\eqref{reparam:Phi1}. That rephasing transformation freedom is particularly
important when counting the number of parameters of the mixing
matrix $\tilde{R}_v$ in Eq.\,\eqref{x->x'}. Since $\tilde{R}_v$ appears in the
couplings involving the physical charged fields it may seem that it is a
physical rotation matrix, needing three angles for its parametrization.
However, only two angles are physical. The reason is that the
reminiscent reparametrization freedom induced by Eq.\,\eqref{reparam:Phi1} can
remove one angle. Such reparametrization freedom is equivalent to rotations
around $\aver{\br}$.
An explicit parametrization using two angles is available in Eqs.\,\eqref{Uv}
and \eqref{Rv} of appendix \ref{ap:V34}.

The case of CP conserving potentials includes the MSSM 2HDM potential (see
Ref.\,\onlinecite{randall:2hdm}) and can be easily analyzed by setting
$\Lambda_{2i}=0$ for $i\neq 2$ and $M_2=0$ (or real $Y$). In addition, if there
is no SPCV, we have $\aver{r_2}=0$. In this case, from Eq.\,\eqref{Mn:gen}, we
see the neutral scalar $t_2$ does not mix with other scalars and corresponds to
a CP-odd field with mass
\eq{
\label{m2}
m^2_2=m^2+\frac{v^2}{4}\Lambda_{22}\,.
}
Relation \eqref{m2} is equivalent to a known relation encountered in the MSSM
[see Eq.\,(10) of Ref.\,\onlinecite{carena}], where the pseudoscalar $t_2$ is
usually called $A$ and $\frac{v^2}{4}\Lambda_{22}= -m^2_W$.
The remaining neutral scalars $t_1,t_3$ are CP-even and their mass matrix can be
also read from Eq.\,\eqref{Mn:gen}.

It is important to stress that the original basis transformations valid before
EWSB forming the $SU(2)_H$ group could be explicitly transposed to the fields
after EWSB.
Although the possibility of transposition could be foreseen, various properties
of the transformations after EWSB could not be anticipated. For example, the
basis transformations after EWSB mix the physical charged Higgs $h^+$ with the
charged Goldstone while the neutral fields $t_i$ mix among them [through the
same $SU(2)_H\sim SO(3)_H$] without mixing with the neutral Goldstone that
transforms as a scalar of $SU(2)_H$.

From the discussions of Sec.\,\ref{sec:Pbasis}, we can see there is an important
distinction between basis transformation and reparametrization. The
transformations of Eq.\,\eqref{uw:trans:gen} constitute legitimate basis
transformations that preserve the gauge structure after EWSB but they do not
configure as reparametrization transformations. On the one hand, only the
original $SU(2)_H$ basis transformations that preserves the $SU(2)_L$ gauge
structure configure as reparametrization transformations. 
On the other hand, the maximal semisimple group of transformations
which mix four real fields, $t_i$ and $G^0$, is $SO(4)$.
In addition,  if we do not impose the kinetic part to be invariant, the
reparametrization group $SU(2)_H$ can be extended to
$SL(2,c)$\,\cite{ivanov:mink}.

As an terminological issue, the term basis transformations (or horizontal
transformations) should be accompanied by the gauge structure that they preserve
to be precise. For example, for the 2HDM treated here, it is important to
specify if the basis transformations act before [$SU(2)_L\otimes U(1)_Y$] or
after EWSB [$U(1)_{\rm EM}$].

In general, the horizontal group after SSB will be larger than the
horizontal group before SSB. It should be remarked that usually the physical
mixing parameters belong to the additional basis transformations only allowed
after SSB. For example, the CKM matrix for quarks comes from the difference
between the rotations on the fields $\{u_L,c_L,t_L\}$ and $\{d_L,s_L,b_L\}$
necessary to diagonalize the respective mass matrices; applying basis
transformations before EWSB, it is only possible to diagonalize one of the up or
down quark Yukawa coupling matrices. A similar structure appears in 2HDMs for
which the mixing among neutral scalars, the matrix $\tilde{R}_v$, appears as the
difference between the PCH basis and the PNH basis.

For general $N$-Higgs-doublet models (NHDMs), the covariant relation for the
mass matrix of neutral scalars can be easily written by generalizing
Eqs.\,\eqref{Mn:gen} and \eqref{w->ti} to $N$ Higgs doublets.
The covariant relation for the mass matrix of charged scalars $\aver{\bbM}$ was
found in Ref.\,\cite{nhdm:V}.
The fields $t_i$, however, will transform as a vector of adj$SU(N)_H$, living
in a real vector space of $N^2-1$ dimensions. Since, in general, a
transformation in $SO(N^2-1)$, a larger group than adj$SU(N)_H$, will be
required to diagonalize $\Mn$, the Physical Neutral Higgs (PNH) basis can not be
reached by reparametrization but only by general horizontal transformations 
valid after EWSB. The corresponding basis transformation group will be
$SO(N^2-1)\otimes SU(N)$, the first factor acting on the neutral scalars and the
second on the charged scalars independently.
The enlargement of the basis transformation group after EWSB compared to
the basis transformation group before EWSB is greater in NHDMs, with $N>2$,
than in the two-Higgs doublet case ($N=2$). But the difference is not just
quantitative. For the 2HDM potential, the basis transformation
group after EWSB, $SU(2)\otimes SU(2)$, is just the double of the basis
transformation group $SU(2)_H$ before EWSB, which can be understood as the
original basis transformation acting independently on the upper $u$ and lower
$w$ components of the doublets, as described in Eq.\,\eqref{uw:trans:gen}.
For $N>2$, the factor $SO(N^2-1)$ necessary to diagonalize $\Mn$ and,
consequently, necessary to reach the Physical basis, can not be thought
as the original reparametrization group $SU(N)_H$ acting independently on the
lower components $w$ of the doublets.

Finally, we can say that a nontrivial horizontal structure in the scalar sector
of a theory enriches the latter significantly, opening the possibility of
different phenomenology such as different symmetry breaking patterns. At the
same time, the theory becomes less predictive as much more free parameters are
available. Nevertheless, useful physical information can be extracted from the
horizontal structure by classifying the transformation properties of the
parameters appearing in the potential. These properties constrain the relations
between parameters before and after SSB, relating, for instance, vectors of the
horizontal group with vectors. In the 2HDM potential analyzed here, we could
relate, for example, the rank-2 tensor $\tilde{\Lambda}$, appearing before
EWSB, with the mass matrix of the neutral scalars $\Mn$, only extractable after
EWSB. Moreover, these relations were basis invariant. It is important to notice
that the transformation properties of the parameters refer to the horizontal
group $SU(2)_H$ acting on the Higgs doublets before EWSB. Although the
horizontal group acting on the fields after EWSB could be larger, the
transformation properties of the parameters followed essentially from the
original horizontal group valid before EWSB. Obviously, a transformation in the
enlarged horizontal group is usually necessary to reach the Physical basis where
all fields have definite masses.

%%%%%%%%%%%%%%%%%%%%%%%%%%%%%%%%%%%%%%%%%%%
\acknowledgments
The author would like to thank Igor Ivanov for usefull discussions.
This work was supported by {\em Fundação de Amparo à Pesquisa do Estado de São
Paulo} (Fapesp).

\appendix
%%%%%%%%%%%%%%%%%%%%%%%%%%%%%%%%%%%%%%%%%%%%%%%%%
\section{Extremum equations}
\label{ap:min}

Any neutral extremum $\aver{r_\mu}$ of the potential in Eq.\,\eqref{V:2+4}
should satisfy the following extremum equations\,\cite{nhdm:V}
\eqarr{
M_0+\Lambda_{00}\aver{r_0}+\bs{\Lambda_0}\sp\aver{\br}=m^2\,,\\
\label{min:i}
M_i+\Lambda_{i0}\aver{r_0}+\tilde{\Lambda}_{ij}\aver{r_j}
=-m^2\aver{\hat{r}_i}\,,
}
where $i=1,2,3$ and $m^2$ is the mass squared of the physical charged Higgs.
The minus sign on the righthand side of Eq.\,\eqref{min:i} is the only
reminiscent of the Minkowski metric adopted in Ref.\,\onlinecite{nhdm:V}.

The original extremum equation on the doublets reads
\eq{
\label{M.w}
\aver{\bbM}\aver{w}=0\,,
}
for $\aver{u}=0$ and $\aver{w}\neq 0$.
Equation \eqref{M.w} means $\aver{w}$ is an eigenvector of $\aver{\bbM}$ with
null eigenvalue.

%%%%%%%%%%%%%%%%%%%%%%%%%%%%%%%%%%%%%%%%%%%%%%%%%%%%%%%%%%%%%5
\section{Bounded below condition}
\label{ap:BBcond}

We seek here the necessary conditions for a bounded below potential using a
method distinct to the ones adopted in
Refs.\,\onlinecite{ivanov:mink} and \onlinecite{maniatis:1}. We will restrict
ourselves to positive definite $V_4$.

Rewriting $V_4$ for $r_0=|\br|$ we obtain
\eq{
\label{V4:ext}
V_4=\mn{\frac{1}{2}}\br^{\tp}(\tilde{\Lambda}+\Lambda_{00}\id_3)\br
+|\br|\br\sp\bs{\Lambda_0}
\,.
}
All variables $r_i$ will be treated here as c-numbers.
We seek the direction $\br$ for which the potential increases more
slowly. We minimize then 
\eq{
V_4'=V_4 + \mn{\frac{1}{2}}\lambda (\br^2-1)\,,
}
constraining $\br$ to be in the unit sphere using the Lagrange multiplier
method.

Differentiating,
\eqarr{
\label{dV:r}
\frac{\partial V_4'}{\partial r_i}&=&
\tilde{\Lambda}_{ij}r_j+\Lambda_{00}r_i
+\Lambda_{0i}|\br|+\hat{r}_i(\br\sp\bs{\Lambda_0})
+ \lambda r_i
\,,~~\\
\label{dV:lambda}
\frac{\partial V_4'}{\partial \lambda}&=&\mn{\frac{1}{2}}(\br^2-1)
\,.
}
Equation \eqref{dV:r} yields
\eq{
\label{BB:r}
\hat{\br}=
-\big[\tilde{\Lambda}+
(\Lambda_{00}+\lambda+\hat{\br}\sp\bs{\Lambda_0})\id_3\big]^{-1}
\bs{\Lambda_0}
\,.
}
The values of $\hat{\br}\sp\bs{\Lambda_0}$ corresponding to an
extremum is given by the roots of
\eq{
\label{root:1}
\hat{\br}\sp\bs{\Lambda_0}=f(\hat{\br}\sp\bs{\Lambda_0}+\lambda)\,,
}
constrained by
\eq{
\label{root:2}
\frac{df(x)}{dx}=1\,,
}
for $x=\hat{\br}\sp\bs{\Lambda_0}+\lambda$.
The function $f(x)$ is defined by
\eq{
f(x)\equiv
-\bs{\Lambda_0}^{\tp}\big[\tilde{\Lambda}+(\Lambda_{00}+x)\id_3\big]
^{-1}\bs{\Lambda_0}
\,.
}
Equation \eqref{root:1} is found by projecting Eq.\,\eqref{BB:r} to
$\bs{\Lambda_0}$ while Eq.\,\eqref{root:2} is equivalent to the requirement
$\hat{\br}\sp\hat{\br}=1$. The components of $\hat{\br}$ perpendicular to
$\bs{\Lambda_0}$ can be found from Eq.\,\eqref{BB:r} once
$\hat{\br}\sp\bs{\Lambda_0}$ is known.

For any extremum satisfying Eq.\,\eqref{BB:r} we find for Eq.\,\eqref{V4:ext}
the value
\eq{
V_4\big|_{\rm extremum}= \mn{\frac{1}{2}}\br^2(-\lambda)\big|_{\rm extremum}
\,.
}
We see all the Lagrange multipliers $\lambda$ corresponding to an extremum
should be negative. In particular, the greatest of them should be negative.

\begin{figure}[h]
\begin{center}
% \fbox{
\includegraphics*[width=6cm,height=8.5cm,angle=-90]{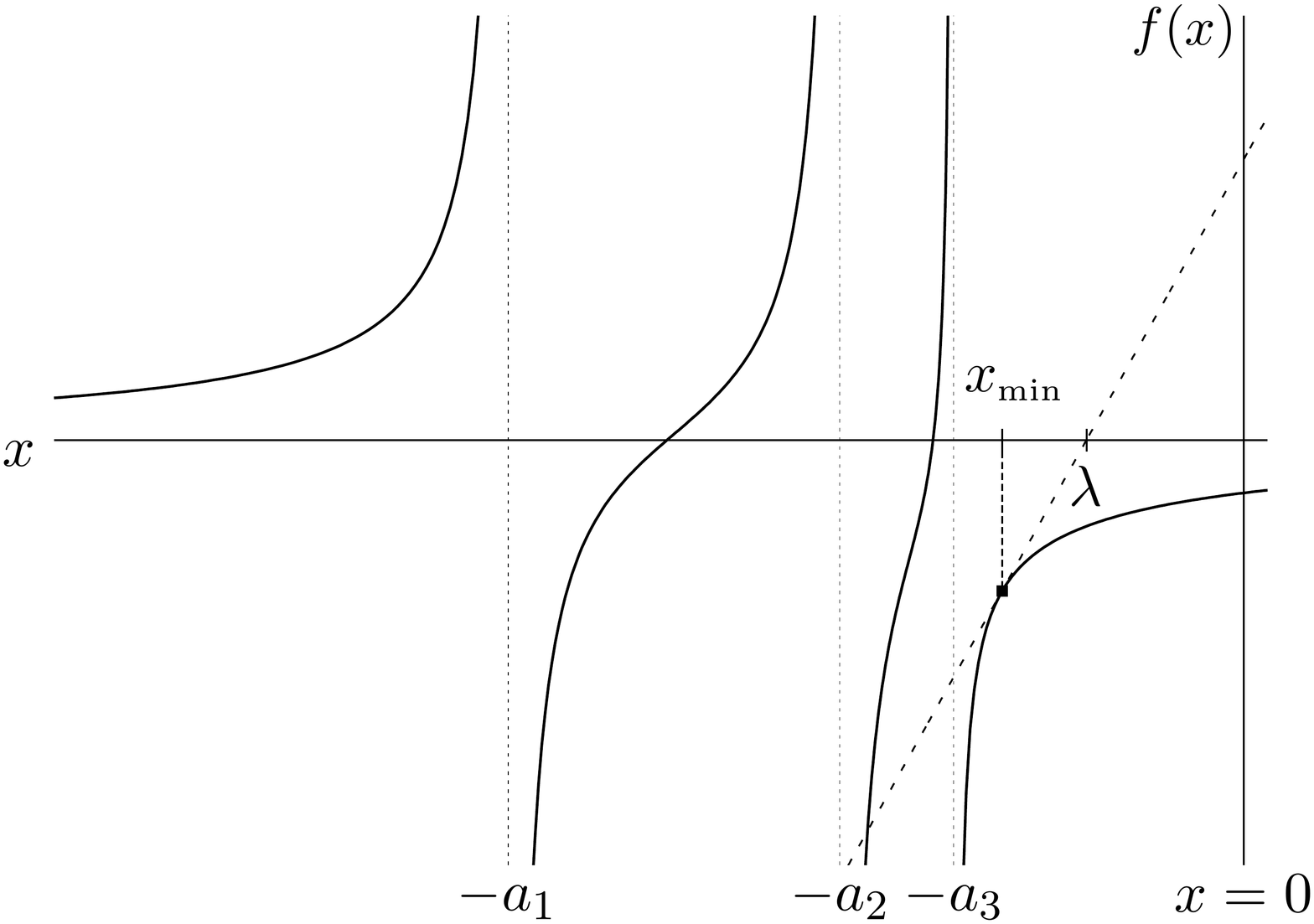}
% \includegraphics*[width=6cm,height=8.5cm,angle=-90]{boundedtex1.ps}
% }
\caption{\label{fig.1}
Plot of typical $f(x)$. The dot lies at $\big(x_{\rm min},f(x_{\rm min})\big)$
where $x_{\rm min}$ is the greatest value that satisfies $f'(x)=1$. Equation
\eqref{root:1} defines the value of $\lambda$ depicted as the intersection of
the line $x-x_{\rm min}+f(x_{\rm min})$ with the $x$-axis.
}
\end{center}
\end{figure}

In the basis for which $\tilde{\Lambda} +\Lambda_{00}\id_3 =
\diag(a_1,a_2,a_3)$, $a_1>a_2>a_3$,
\eq{
f(x)=-\sum_{i=1}^{3}\frac{\Lambda^2_{0i}}{a_i+x}\,.
}
A plot of $f(x)$, with $a_i>0$, can be seen in Fig.\,\ref{fig.1} jointly with
the solution of greatest $\lambda$. We see there are at least two extrema
corresponding to the least and greatest $\lambda$.
The intermediary extrema may not exist depending on the minimum slope of the
curves. For example, in Fig.\,\ref{fig.1}, for $-a_2\le x\le -a_3$, there is no
solution for $f'(x)=1$.

From the schematic view of Fig.\,\ref{fig.1} we see $a_i>0$ is necessary to
have positive $-\lambda$ and consequently positive definite $V_4$, unless
$\Lambda_{0i}=0$ for nonpositive $a_i$. In a general basis, it is necessary that
\eq{
\label{BB:Lt}
\Lambda_{00}+\text{eigenvalues}(\tilde{\Lambda})>0\,,
}
unless $\bs{\Lambda_{0}}$ have null projection in some eigenvector direction.
To assure $V_4$ is positive definite is necessary and sufficient
to have the greatest Lagrange multiplier
\eq{
\label{BB:lagrange}
\max\lambda <0\,.
}
From Eq.\,\eqref{root:1}, the distance between the greatest $x_{\rm min}$ and
the greatest $\lambda$ is
$|\hat{\br}\sp\bs{\Lambda_0}|= -\hat{\br}\sp\bs{\Lambda_0}$.

Finally, if $a_i>0$ and one makes $|\Lambda_{0i}|$ small enough, we can always
find $\lambda<0$, proving the assertion preceding Eq.\,\eqref{BB:L0}. As 
$|\Lambda_{0i}|$ get smaller, the curves of $f(x)$ get closer to the $x$-axis.
In special, from $f(x)\ge f(x)|_{a_1,a_2\rightarrow a_3}$, $a_3=\min(a_i)$, we
can conclude that 
\eq{
\label{cond:suf:1}
2|\bs{\Lambda_0}|< a_3\,
}
is a sufficient condition.

For $|r_0|>|\br|$ we can parametrize $r_0=e^{\chi}|\br|$, $\chi> 0$. The
analysis of the minimization of $V_4$ for fixed $|\br|$ is equivalent to
the preceding analysis replacing $\Lambda_{00}\rightarrow\Lambda_{00}e^{2\chi}$
and $\Lambda_{0i}\rightarrow\Lambda_{0i}e^{\chi}$. If
$|\bs{\Lambda_0}|/\Lambda_{00}\le 1$, condition \eqref{cond:suf:1} is preserved
for $\chi>0$ once it is valid for $\chi=0$. If
$|\bs{\Lambda_0}|/\Lambda_{00}>1$ and $\lambda_i>0$, a sufficient condition is
\eq{
\frac{|\bs{\Lambda_0}|^2}{\Lambda_{00}}<\min(\lambda_i)\,,
}
where $\lambda_i$ are the eigenvalues of $\tilde{\Lambda}$.

%%%%%%%%%%%%%%%%%%%%%%%%%%%%%%%%%%%%%%%%%%%%%%%%%
\section{Proof of Eq.\,$\mbox{\eqref{s0->ti}}$}
\label{ap:s0->ti}

From the completeness of the $\sigma_\mu$ matrices\,\cite{ccn:fierz},
\eq{
\mn{\frac{1}{2}}(\sigma_\mu)_{ab} (\sigma_\mu)_{cd}=
\delta_{ad}\delta_{cb}\,,
}
we can calculate, for a neutral vacuum $\aver{w}\neq 0$,
\eqarr{
\aver{r_0}s_0+\aver{r_i}s_i
&=&
\mn{\frac{1}{4}}
\Tr[\aver{w}\aver{w}^\dag\sigma_\mu]
\Tr[\sigma_\mu(w\aver{w}^\dag+h.c.)]
\cr
&=&
\mn{\frac{1}{2}}
\Tr[(w\aver{w}^\dag+h.c.)\aver{w}\aver{w}^\dag]
\cr
&=&
2\aver{r_0}s_0
\,,
}
where $\aver{r_0}=|\aver{w}|^2/2=v^2/4$.
Hence,
\eq{
s_0=\aver{\hat{r}_i}s_i\,,
}
since $\aver{r_0}=\sqrt{\aver{r_i}\aver{r_i}}$.
With the same reasoning, one can prove
\eq{
s_0y_0=s_iy_i\,.
}
%%%%%%%%%%%%%%%%%%%%%%%%%%%%%%%%%%%%%%%%%%%%%%%%%%%%%%%%%5
\section{Basis covariance for $\aver{\bbM}$}
\label{ap:covariance:M}

It is important to stress that the definition of the charged mass matrix
$\aver{\bbM}$ is covariant by basis transformation \eqref{Phi->PCH} in the
following sense. The definition of the charged mass matrix in
Eq.\,\eqref{M:charged} is valid in any basis, in particular, in the PCH basis,
\eq{
\diag(m^2,0)=\aver{\bbM'}=Y'+
\mn{\frac{1}{2}}\sigma_\mu\Lambda'_{\mu\nu}\aver{r'_\nu}\,.
}
The equation \eqref{bbM->diag} can thus be written as
\eq{
U^\dag_v\big(Y+\mn{\frac{1}{2}}\sigma_\mu\Lambda_{\mu\nu}\aver{r_\nu}\big)U_v
=
Y'+
\mn{\frac{1}{2}}\sigma_\mu\Lambda'_{\mu\nu}\aver{r'_\nu}
\,,
}
where the relation between $\{Y',\Lambda',\aver{r'}\}$ in the PCH basis and
$\{Y,\Lambda,\aver{r}\}$ in the original basis is
\eqarr{
Y'&=& U^\dag_v Y U_v\,,\\
\aver{r'_\mu}&=& R^{\tp}_{\mu\nu}\aver{r_\nu}\,,\\
\Lambda'_{\mu\nu}&=& R^{\tp}_{\mu\alpha}\Lambda_{\alpha\beta}R_{\beta\nu}
\,,
}
and $R_{\mu\nu}=R_{\mu\nu}(U_v)$ is given by Eq.\,\eqref{R.munu}.
Hence, the first term of \eqref{V2:SSB:uw} is form invariant,
\eq{
u^\dag\aver{\bbM}u=u'^\dag\aver{\bbM'}u'\,.
}

%%%%%%%%%%%%%%%%%%%%%%%%%%%%%%%%%%%%%%%%%%%
\section{Interaction terms}
\label{ap:V34}
\def\bx{\mathbf{x}}
\def\by{\mathbf{y}}
\def\bt{\mathbf{t}}

The interaction terms can be simplified into
\eqarr{
\frac{v}{2}V_3\big|_{\mt{\rm SSB}}&=&
-\big[u^\dag Y u + w^\dag Yw\big]t_{\parallel}
+m^2[x_0\,t_{\parallel}-\bx\sp\bt]~~
\cr&&
+\aver{r_0}\bs{\Lambda_0}_{\perp}\sp\bt_{\perp}\,\big[
|u_{\perp}|^2+\mn{\frac{1}{2}}\bt^2_{\perp}
\big]
\cr&&
+(\bx+\by)^{\tp}\!\Mn\,\bt
% \cr&&
\,,
}
\eqarr{
V_4\big|_{\mt{\rm SSB}}\!\!&=&
\mn{\frac{1}{2}}\Lambda_{00}(r^2_0-\br^2)
+(r_0-\aver{\hat{\br}}\sp\br)\bs{\Lambda_0}\sp\br
\cr&&
+~\frac{\br^{\tp}\!\Mn\br}{2\aver{r_0}}
% \cr&&
+~\mn{\frac{1}{2}}\big(\Lambda_{00}-\frac{m^2}{\aver{r_0}}\big)
% \br^{\tp}\big[\id_3-\!\aver{\hat{\br}}\!\aver{\hat{\br}}^{\!\tp}\!\big]\br
\br_{\perp}^2
\,,~
}
where
\eqarr{
u_{\perp}&\equiv& u-\aver{\hat{w}}\aver{\hat{w}}^\dag u\,,\\
t_{\parallel}&\equiv& \bt\sp\aver{\hat{\br}}\,,\\
\bt_{\perp}&\equiv&\bt-\aver{\hat{\br}}t_{\parallel}\,,\\
|u_{\perp}|^2&=&x_0-\bx\sp\aver{\hat{\br}}\,,\\
\bt_{\perp}^2&=&2(y_0-\by\sp\aver{\hat{\br}})
\,,
}
and $\bs{\Lambda_{0}}_{\perp}$ is analogous to $\bt_{\perp}$.

One can also explicit the matrices $U_v$ and $\tilde{R}(U_v)$ in
Eqs.\,\eqref{u->u'} and \eqref{x->x'} choosing an explicit parametrization:
\eqarr{
\label{Uv}
U_v&=&U(\theta_v,\xi)i\sigma_2\,,\\
\label{Rv}
\tilde{R}_v&=&\tilde{R}(\theta_v,\xi)\tilde{R}(\pi,0)
=\tilde{R}(U_v)
}
where
\eqarr{
\label{U:thetaphi}
U(\theta,\varphi)&\equiv&
e^{-i\frac{1}{2}\sigma_3\varphi}
e^{-i\frac{1}{2}\sigma_2\theta}
\,.\\
\label{R:thetaphi}
\tilde{R}(\theta,\varphi)&\equiv&
e^{-iJ_3\varphi}
e^{-iJ_2\theta}\,.
}
The generators of rotations are defined by $i(J_k)_{ij}=\epsilon_{ijk}$.

%%%%%%%%%%%%%%%%%%%%%%%%%%%%%%%%%%%%%%%%%%%

%%%%%%%%%%%%%%%%%%%%%%%%%%%%%%%%%%%%%%%%%%%%%%%%%

\begin{thebibliography}{99}
\bibitem{endnote-1}
A set of continuous minima (of different orbits) may exist in certain symmetric
potentials\,\cite{ivanov:mink2}.
\bibitem{endnote0}
The terminology ``horizontal transformations'' and ``basis transformations'',
as well as ``horizontal group'' and ``basis transformation group'', will be
used interchangeably.
\bibitem{endnote1}
The mass squared of the physical charged scalar $h^+$, $m^2$, can be regarded as
a scalar under $SU(2)_H$ by defining $m^2=\Tr[\aver{\bbM}]$.
\end{thebibliography}
\end{document}